\documentclass[aps,prl,reprint]{revtex4-1}
\usepackage{amsmath}
\usepackage{graphicx}
\usepackage{color}
\usepackage{amssymb}
\usepackage{wasysym}
\usepackage{amsthm}
\usepackage{soul}
\usepackage{mathtools}   % loads �amsmath�
\usepackage[T1]{fontenc}
\usepackage[english]{babel}
\usepackage{hyphenat}
\input{epsf}

\begin{document}

\bibliographystyle{apsrev}

\renewcommand{\andname}{\ignorespaces}

\title{Athermal Fracture of Elastic Networks:\\ How Rigidity Challenges the Unavoidable Size-Induced Brittleness}

\author{Simone Dussi}
\email{simone.dussi@wur.nl}
\author{Justin Tauber}
\author{Jasper van der Gucht}
\affiliation{Physical Chemistry and Soft Matter, Wageningen University, Stippeneng 4, 6708 WE, Wageningen, The Netherlands}

\begin{abstract}
  By performing extensive simulations with unprecedentedly large system sizes, we unveil how rigidity influences the fracture of disordered materials. We observe the largest damage in networks with connectivity close to the isostatic point and when the rupture thresholds are small. However, irrespectively of network and spring properties, a more brittle fracture is observed upon increasing system size. Differently from most of the fracture descriptors, the maximum stress drop, a proxy for brittleness, displays a universal non-monotonic dependence on system size. Based on this uncommon trend it is possible to identify the characteristic system size $L^*$ at which brittleness kicks in. The more the disorder in network connectivity or in spring thresholds, the larger $L^*$. Finally, we speculate how this size-induced brittleness is influenced by thermal fluctuations.
\end{abstract}

\maketitle

\noindent
Since the pioneering work of Griffith on crack nucleation in ordered materials with isolated defects~\cite{griffith}, the last 30 years have witnessed a still-growing interest for fracture occurring in intrinsically disordered materials~\cite{sahimi1986,duxbury1987,kahng1988,zapperi1997,alava2006,moreira2012,shekhawat2013,herrmann_book2014}. 
The fracture behavior of a material is intimately related to its microscopic structure, especially when thermal fluctuations are not relevant. How stress is (re)distributed during deformation and after bond-breaking events determines the material mechanical response. Abruptness is the hallmark of brittle fracture: the mechanical response of a brittle material subjected to an increasing strain deformation suddenly vanishes after reaching a peak. This is the macroscopic consequence of stress concentration in the system that induces bond-breaking events and triggers the nucleation of a crack that irremediably propagates throughout the system. For (more) ductile systems, such catastrophic breaking event is absent or postponed much after the stress peak. As a consequence, the mechanical response of ductile materials persists after the peak. 

\begin{figure}[h!t]
	\begin{center}
		\includegraphics[width=0.49\textwidth]{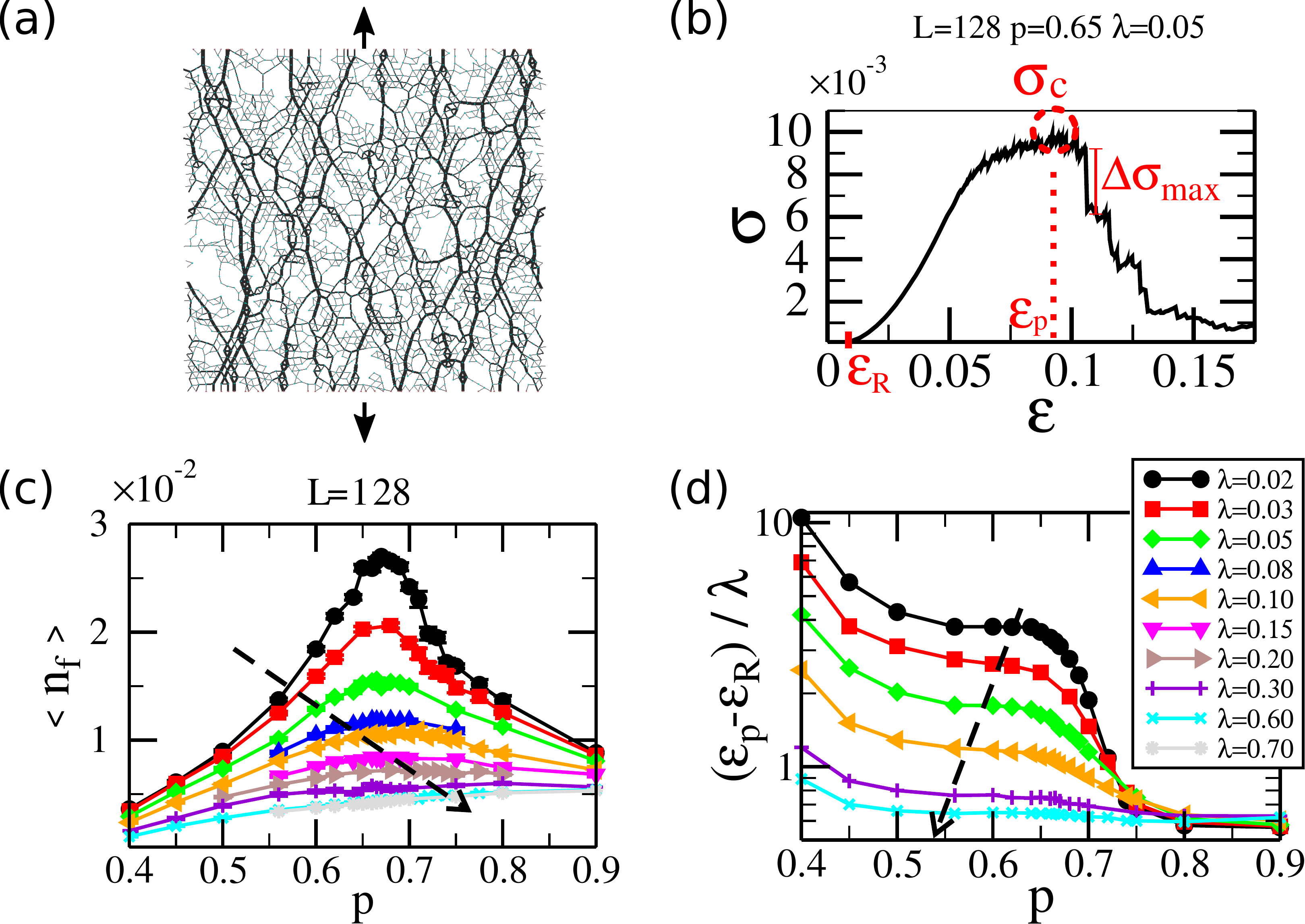}
	\end{center}
	\caption{Fracture depends on connectivity and thresholds. (a) When deforming diluted elastic networks, aligned sets of load-bearing bonds, called force chains, appear. Line thickness quantifies spring deformation. (b) Example of stress $\sigma$ - strain $\epsilon$ curve and fracture descriptors. (c) Fraction of broken bonds at failure $n_{\text{f}}$ and (d) \emph{hidden length} exploited in the fracture process, as a function of the connectivity parameter $p$ for different rupture threshold $\lambda$ and fixed system size $L=128$.}
	\label{fig1}
\end{figure}

Biopolymer networks, such as collagen, are ubiquitous examples of disordered structures, with a peculiarly low connectivity that places these materials below the isostatic point of mechanical stability~\cite{maxwell}. At small deformation, these networks would not be rigid without their fiber bending stiffness, responsible for their small linear modulus. However, exactly because of their sub-isostaticity these materials show an extraordinary strain-stiffening, i.e. their modulus increases by orders of magnitude when deformed above an onset strain. This mechanical response is a consequence of an athermal strain-driven rigidity transition: beyond the onset strain, the stretching of the bonds starts to control the network response and is sufficient to rigidify the system~\cite{onck2005,wyart2008,licup2015,sharma2016,feng2016,broedersz_review,vermeulen2017,burla2019,shivers2019a}. In fact, when deformed, these diluted elastic networks exhibit a very heterogeneous stress distribution with emerging force chains~\cite{heussinger2007,arevalo2015,liang2016,zhang2017,shivers2019b} (see Fig.~\ref{fig1}(a)), similarly to deformed granular and porous materials~\cite{cates1998,ohern2001,majmudar2005,laubie2017}. The effects of these heterogeneous stresses on fracture have been only partially addressed. 

It was recently argued that a continuous breakage and formation of force chains in sub-isostatic networks leads to a complete suppression of stress concentration, thereby preventing crack nucleation at all length-scales~\cite{zhang2017}. This conclusion clashes with a recent theory (that does not explicitly account for material rigidity) that predicts crack nucleation for any disordered system approaching the thermodynamic limit (infinite system size)~\cite{shekhawat2013}. This raises the question whether sparse elastic networks show fracture behavior that differs qualitatively from that of other disordered materials. Evidence based on simulations of moderate system size~\cite{driscoll2016,zhang2017} and fracture experiments on small meta-materials~\cite{driscoll2016,berthier2019,qin2017} shows that rigidity cannot be neglected if one wishes to understand fracture in these materials. More generally, despite recent progress~\cite{ellenbroek2015,malakhovsky2006,girard2012,laubie2017,bouzid2017,bonfanti2018}, a clear link between network structure and fracture is still missing, even for a cornerstone such as the central-force spring network model. In this Letter, by fully characterizing the fracture of such a model, we aim to demonstrate whether the tuning of rigidity can actually suppress crack nucleation or not.

We perform (off-lattice) simulations of diluted spring networks with different connectivity, rupture thresholds, topologies, and unprecedentedly large system sizes. Here, we focus on triangular networks made of $L \times L$ nodes in which a fraction $1-p$ of the bonds is removed. Results from other topologies are reported in the Supplemental Material~\cite{SM}. All bonds are harmonic springs with unit stiffness and unit rest length. A bond breaks irreversibly when its deformation exceeds the rupture threshold $\lambda$ (same for all springs). We deform the networks uniaxially in the $y$-direction under athermal and quasistatic loading conditions, by applying small strain steps and using the FIRE algorithm~\cite{bitzek2006} to minimize the energy in between strain steps and breaking events. We characterize the network response by measuring the stress $\sigma$, defined as the $yy$-component of the virial stress tensor, as a function of the (engineering) strain $\epsilon =  (L_y-L_y^{(0)})/L_y^{(0)}$, where $L_y$ and $L_y^{(0)}$ are the system dimension during and before the deformation, respectively. We identify the strain $\epsilon_{\text{R}}$ at which the network rigidifies (i.e. when $\sigma$ exceeds the numerical noise), the maximum stress $\sigma_{\text{c}}$ and its associated strain value $\epsilon_{\text{p}}$, and the maximum stress drop $\Delta \sigma_{\text{max}}$, as illustrated in Fig.~\ref{fig1}(b). Furthermore, we calculate the fraction of broken bonds at final failure (defined when the network is broken in two parts) $n_{\text{f}}=N_{\text{f}}/N_{\text{in}}$, with $N_{\text{f}}$ the number of broken bonds at failure and $N_{\text{in}}$ the number of intact bonds at rest. Quantities are expressed in reduced units and averaged over many configurations (see Supplemental Material~\cite{SM} for details).

In Fig.~\ref{fig1}, we show results at fixed system size $L=128$. In panel (c), we observe for small $\lambda$ a non-monotonic dependence of $n_{\text{f}}$ on the network connectivity, with a maximum around the isostatic point~\cite{maxwell} ($p_{\text{iso}}\simeq0.66$), consistently with Ref.~\cite{zhang2017}. Approaching the limit of a fully connected network ($p=1$), the number of broken bonds drastically diminishes, as expected for a more homogeneous material that typically breaks in a brittle fashion via crack nucleation at the weakest spot followed by a localized propagation involving only few bonds. Also at the other extreme of the connectivity range, close to the geometric percolation limit $p_{\text{g}}\simeq 0.347$, only few bonds are needed to break the already loosely connected networks. Interestingly, when increasing $\lambda$, that is the strength of the individual springs, $n_{\text{f}}$ decreases for all $p$ and the maximum around $p_{iso}$ disappears. Therefore, a general gradual transition to a more localized type of fracture is observed when the breaking process starts at larger deformation (larger $\lambda$) that corresponds to a larger system response (higher stress and modulus). This implies that a more rigid system breaks in a more brittle fashion. To further quantify the role of the network architecture on the fracture process, we plot in Fig.~\ref{fig1}(d) the hidden length emerging during the deformation, defined as the strain interval $\epsilon_{\text{p}}-\epsilon_{\text{R}}$ in which the (rigidified) network reaches the stress peak, normalized by the stretchability of the individual elements $\lambda$. For small $\lambda$, we observe that $(\epsilon_p-\epsilon_R)/\lambda > 1$ for networks with $p \lessapprox p_{iso}$, meaning that they can be stretched significantly more than their elements before network fracture occurs. This is a macroscopic consequence of the underlying complex and heterogeneous stress (re)distribution during the deformation, involving non-affine displacements and formation/breakage of force chains~\cite{heussinger2007,broedersz_review,zhang2017,ellenbroek2009}. By contrast, for large $p$ and/or for large $\lambda$ the hidden length is less than unity, indicating that significant stress concentration occurs before exploiting all the possible bonds stretchability. From our analysis (other quantities are shown in the Supplemental Material~\cite{SM}) we can evidently conclude that the fracture process can be tuned by varying $p$ and $\lambda$ from brittle (abrupt post-peak response, fracture is localized in space and time) to more ductile (more continuous post-peak response, and damage is larger and more diffuse) when $L=128$. We confirmed the universality of the observed behavior by simulating two- and three-dimensional diluted networks with different topologies (see Supplemental Material~\cite{SM}).

\begin{figure}[h!t]
\begin{center}
\includegraphics[width=0.49\textwidth]{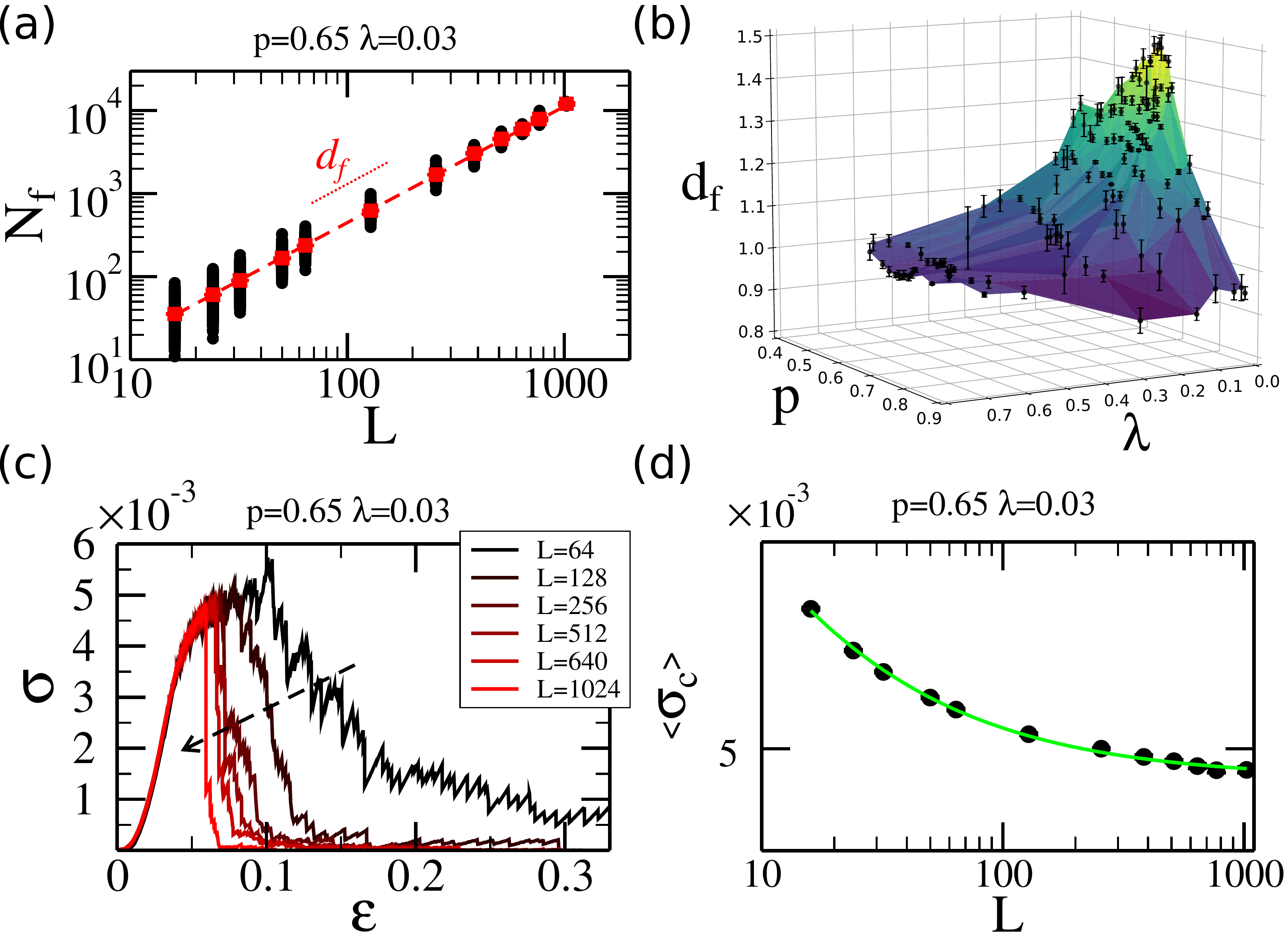}
\end{center}
\caption{The role of system size. (a) Size scaling of number of broken bonds $N_{\text{f}}$ for triangular networks with $p=0.65$, $\lambda=0.03$. Black circles indicate single-run values, red squares are averages and the dashed line is the power-law fit used to extract $d_f$. (b) Damage fractal dimension $d_{\text{f}}$ as a function of $p$ and $\lambda$. (c) Single-run stress-strain curve for increasing system size $L$, showing that response is abrupt for large $L$. (d) Size scaling of the maximum stress $\sigma_{\text{c}}$ (log-log plot). Symbols correspond to averages and the line is the best fit. Data follow a power-law decay.}
\label{fig2}
\end{figure}

\begin{figure*}[h!t]
\begin{center}
\includegraphics[width=0.80\textwidth]{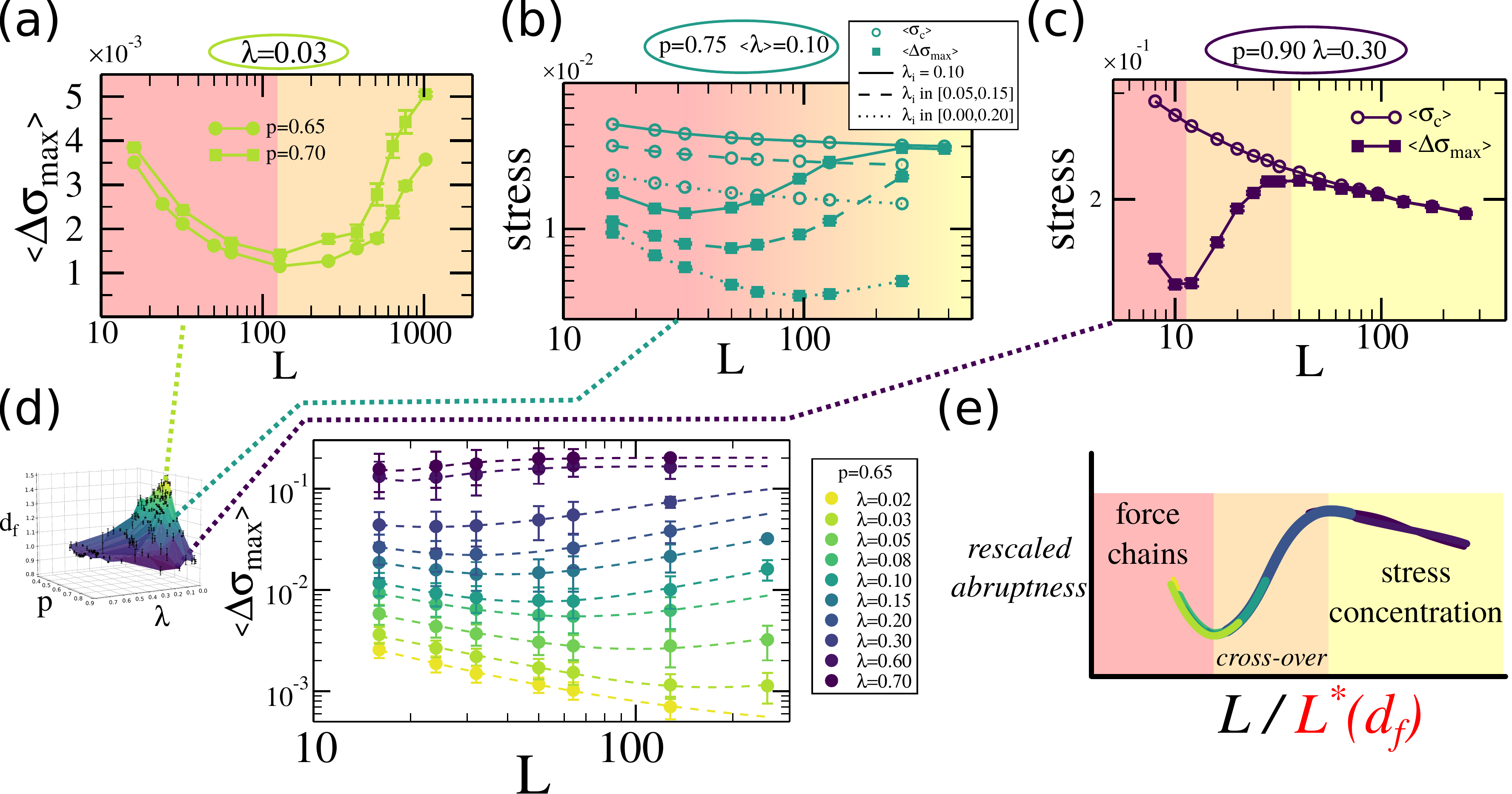}
\end{center}
\caption{Regimes based on fracture abruptness. (a) Size scaling of fracture abruptness $\Delta \sigma_{\text{max}}$ for networks close to the isostatic point with small $\lambda$. A minimum in the trend is evident. (b) Size scaling of maximum stress (empty circles) and fracture abruptness (filled squares) for diluted networks ($p=0.75$) with all springs with same $\lambda_i=0.10$ (solid lines), and with additional disorder in the spring thresholds $\lambda_{\text{i}}$ uniformly distributed in the two intervals indicated in the legend (dashed and dotted lines). The additional threshold disorder shifts the minimum to larger $L$. (c) For $p=0.90$ and $\lambda=0.30$, a third region is probed for large $L$ where the abruptness scales as the maximum stress. (d) Size scaling of abruptness close to the isostatic point for different $\lambda$, color-coded with the associated damage fractal dimension $d_{\text{f}}$ (inset). (e) Universal trend for rescaled abruptness (see also Supplemental Material~\cite{SM}): at small $L$ force chains dominate the mechanical response but beyond $L^*$ stress concentration leading to abrupt fracture is unavoidable.}
\label{fig3}
\end{figure*}

Next, we study the effect of system size, known to be crucial in fracture~\cite{alava2006,alava2008,arevalo2015,taloni2018}, to understand if the locally inhomogeneous stress (re)distribution still has macroscopic implications in larger systems. First, we consider the total number of broken bonds at failure $N_{\text{f}}$ as a function of the system size $L$ and extract the damage fractal dimension $d_{\text{f}}$ by fitting $N_{\text{f}} \sim L^{d_{\text{f}}}$, as shown in Fig.~\ref{fig2}(a) for $p=0.65$, $\lambda=0.03$. In Fig.~\ref{fig2}(b), we plot $d_{\text{f}}$ for several connectivities and thresholds. We observe that $d_{\text{f}}$ is larger close to the isostatic point and for small $\lambda$. When increasing $\lambda$ and/or moving away from $p_{\text{iso}}$, $d_{\text{f}}$ gradually decreases and approaches the expected $d_{\text{f}}=1$ for brittle fracture where a crack propagates almost in a straight line, and therefore the number of broken bonds scales linearly with system size. The upper bound of $d_{\text{f}}=2$ for which damage is completely delocalized~\cite{moreira2012} might be expected only when $\lambda \rightarrow 0$. Secondly, in Fig.~\ref{fig2}(c), we show the stress-strain curve for different system sizes of networks close to the isostatic point and with small $\lambda$. Despite the fact that $d_{\text{f}}$ is maximal for these networks, we observe that their response becomes evidently more brittle for larger $L$. We also note that the maximum stress $\sigma_{\text{c}}$, as exemplary shown in Fig.~\ref{fig2}(d), decreases upon increasing system size, as commonly observed in fracture studies, towards a limiting value $\sigma_{\text{c}}^{(\infty)} \neq 0$ that depends on $p$ and $\lambda$. On passing, we note that our data suggest that $\sigma_c$ decays following a power-law, therefore in a qualitatively different way compared to the scalar models that have been extensively used so far to describe fracture of disordered materials~\cite{duxbury1987,manzato2012,roy2017b} (see also Supplemental Material~\cite{SM}). Finally, we quantify the fracture abruptness by looking at the maximum stress drop $\Delta \sigma_{\text{max}}$. Strikingly, instead of a conventional monotonic decay with increasing system size, we observe a non-monotonic trend, as shown in Fig.~\ref{fig3}(a) for networks close to $p_{\text{iso}}$ and with $\lambda=0.03$. For small $L$, $\Delta \sigma_{\text{max}}$ decreases upon increasing system size. For a second-order transition, for which the system response vanishes continuously, we would expect $\Delta \sigma_{\text{max}} \rightarrow 0$ for $L \rightarrow \infty$. However, the decrease in abruptness stops around $L \simeq 128$. At the same time, the increase of $\Delta \sigma_{\text{max}}$ observed for networks containing up to a few million bonds ($L=1024$) cannot continue for even larger (computationally inaccessible) system sizes since $\Delta \sigma_{\text{max}} \leq \sigma_{\text{c}}$ must hold and we found that $\sigma_{\text{c}}$ decreases with increasing $L$. Indeed, when simulating networks with $p=0.75$ and $\lambda=0.10$, as shown in Fig.~\ref{fig3}(b) (solid lines), we observe a slower increase of $\Delta \sigma_{\text{max}}$ when approaching the limiting value of $\sigma_{\text{c}}$. A third region where $\Delta \sigma_{\text{max}} \sim \sigma_{\text{c}}$, indicative of brittle fracture, becomes clear in panel (c) where we show results for $p=0.90$ and $\lambda=0.30$. Therefore, there must exist a relation between the rigidity-controlled damage, quantified via $d_f$, and the system size intervals corresponding to the three fracture regimes. Indeed, when plotting in panel (d) the abruptness for simulations at fixed $p=0.65$ and increasing $\lambda$, i.e. decreasing the associated $d_{\text{f}}$, it seems that each individual curve represents a different part of a universal response. In fact, we were able to manually rescale and collapse the data onto a master curve (see Supplemental Material~\cite{SM}), schematically depicted in panel (e). At present, we are unable to provide an analytical expression for the scaling function. Nevertheless, the transitions between the three regimes must reflect the underlying stress (re)distribution processes. For small $L$, the mechanical response is dominated by breakage and reformation of force chains and $\Delta \sigma_{\text{max}}$ initially decreases when more force chains are present upon increasing $L$. However, above a certain length-scale $L^*$ (either the maximum or the minimum of the curve) force chains are ineffective in avoiding stress concentration and brittle fracture is observed. Both network structure and spring properties control these transitions. To further prove this statement, we consider additional disorder by drawing random thresholds from a uniform distribution. In Fig.~\ref{fig3}(b), we show results for networks with the same $p=0.75$ and average $\langle \lambda \rangle=0.10$, but with individual spring thresholds $\lambda_i$ drawn from two different intervals (as indicated in the legend) and when no disorder is present ($\lambda_i=0.10 \; \forall i$). We observe that the minimum in $\Delta \sigma_{\text{max}}$ shifts to larger $L$ when the threshold distribution is broader. In short, the larger the (connectivity or threshold) disorder, the larger the system size $L^*$ at which brittleness kicks in. We find that the non-monotonic size-scaling of fracture abruptness is universal, irrespectively of network topology or technical details regarding boundary conditions and simulation protocols (see Supplemental Material~\cite{SM}).
 
\begin{figure}[h!t]
\begin{center}
\includegraphics[width=0.49\textwidth]{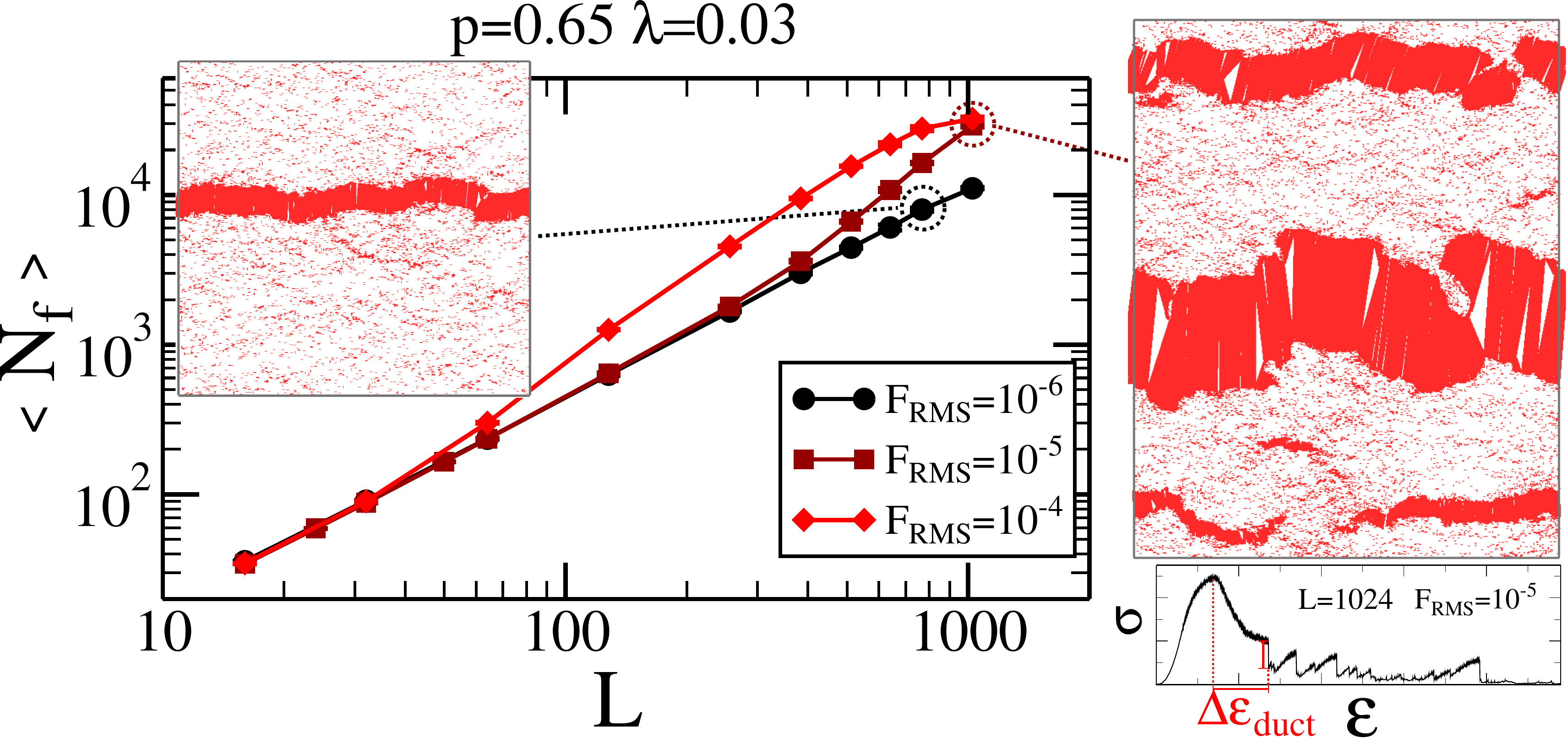}
\end{center}
\caption{Athermal limit and macroscopic cracks. Main graph: size scaling of broken bonds for simulations performed with different energy minimization tolerance $F_{\text{RMS}}$. Larger $L$ requires lower $F_{\text{RMS}}$ to perform simulations in the athermal limit, where fracture occurs due to a single macro-crack (see post-mortem snapshot on the left, where only broken bonds are shown). For larger $F_{\text{RMS}}$ and huge system sizes, multiple macro-cracks are observed (right snapshot) and the network response shows a large ductile interval $\Delta\epsilon_{\text{duct}}$ (bottom right).
}
\label{fig4}
\end{figure}

Our study shows that the fracture process of truly large ($L>L^*$) elastic networks is always dominated by stress concentration leading to unavoidably abrupt fracture when the critical deformation ($\epsilon_p,\sigma_c$) is reached. In contrast with the conclusions of Ref.~\cite{zhang2017}, we have shown that also sub-isostatic networks break via crack nucleation when sufficiently large. Our results are consistent with the idea of crack nucleation as limiting fracture behavior for disordered materials~\cite{shekhawat2013}. However, differently from the finite-size criticality of Ref.~\cite{shekhawat2013} that manifests itself as a smooth cross-over in exponents associated to (avalanche or crack size) distributions, the size-induced brittleness studied here gives rise to a non-monotonic size-dependence of $\Delta \sigma_{\text{max}}$. The extreme point(s) of such a trend can be used to define a characteristic size $L^*$ corresponding to the transition from ductile to brittle crack nucleation. Despite the abrupt nature of fracture, we have also shown that the damage can be delocalized in space due to the inhomogeneous network structure. In fact, the damage zone spans the entire system with a distinct fractal dimension $d_{\text{f}}>1$, as illustrated in the left post-mortem snapshot of Fig.~\ref{fig4}. This is reminiscent of some fracture modes investigated using fiber bundle models~\cite{roy2017}, where however the range of stress redistribution is an imposed parameter instead of an emerging structure property.

Finally, we conclude this Letter by showing an unexpected dependence on the tolerance parameter $F_{\text{RMS}}$ used in the energy minimization protocol employed in our simulations.~\cite{bitzek2006,SM} $F_{\text{RMS}}$ is the maximum force per node allowed in the system after a deformation step or breaking event. Typically, $F_{\text{RMS}}$ is chosen small enough to ensure that the system is in its energy minimum, or equivalently in mechanical equilibrium, and therefore the simulation is in the athermal (energy-dominated) limit. In Fig.~\ref{fig4}, we plot the size scaling of $N_{\text{f}}$ for different $F_{\text{RMS}}$ and we observe that for the huge system sizes investigated here a great number of additional bonds are broken when a larger $F_{\text{RMS}}$ is used. Correspondingly, a more ductile response is obtained due to the formation of multiple macro-cracks. Therefore, we remark that (i) to reach the athermal limit when a large $d_{\text{f}}$ is expected, an increasingly smaller $F_{\text{RMS}}$ might be needed to simulate increasingly larger networks; (ii) since using a larger $F_{\text{RMS}}$ corresponds to push the system close but not at its energy minimum, $F_{\text{RMS}}$ could be a proxy for temperature. According to this interpretation, thermal fluctuations would couple with large rigidity fluctuations around the isostatic point, giving rise to ductile behavior even for large system sizes. Further studies are needed to confirm the latter intriguing speculation.

In summary, our results demonstrate that athermal networks unavoidably break in a brittle fashion when approaching the thermodynamic limit, i.e. when increasing the system size beyond a characteristic length-scale $L^*$. We have shown that $L^*$ can be controlled by tuning individual elements or their assembly, with larger disorder increasing $L^*$. Furthermore, we found that the fractal dimension of the damage zone $d_{\text{f}}$ is coupled with $L^*$ and can be very large close to the isostatic point~\cite{driscoll2016,zhang2017}, which can be interpreted as a critical point for fracture. These considerations are relevant not only for metamaterials (for which $L$ is small and $\lambda$ or the network geometry can be easily tuned) but also for biological samples, since they are also often far from the thermodynamic limit. For example, biopolymer networks between two cells should have sizes well below $L^*$ to prevent catastrophic failure, and even reconstituted collagen networks inside a rheometer have shown size-dependent fracture behavior~\cite{arevalo2010}.

\smallskip
\emph{Acknowledgments ---} This work is part of the SOFTBREAK project funded by the European Research Council (ERC Consolidator Grant). S.D. thanks Xiaoming Mao for informative correspondence. We thank Jessi van der Hoeven for a critical reading of the manuscript. 

\bibliography{biblio}

\clearpage

\section{Supplemental Material for \\ ``Athermal Fracture of Elastic Networks:\\ How Rigidity Challenges the Unavoidable Size-Induced Brittleness''}

\renewcommand{\thefigure}{S\arabic{figure}}
\noindent

\section{Simulation methods}
We generate networks by diluting regular lattices consisting of $L \times L$ nodes with spacing $l_0$ and different topologies. A fraction $p$ of all possible nearest-neighbor bonds is present. The associated average network connectivity is therefore $\langle z \rangle = p \, z_{\text{max}}$, with $z_{\text{max}}$ depending on network topology (see also Fig.~\ref{SfigB}). All bonds are harmonic springs with stiffness $\mu$ and rest length $l_0$ and break irreversibly when exceeding the threshold $\lambda$ (same for all springs). By setting $l_0=1$ and $\mu=1$, we can express all quantities in reduced units, and $\lambda$ can be equivalently considered as deformation or force threshold. The energy of the system reads $\mathcal{H}=\frac{\mu}{2l_0}\sum_{\langle ij \rangle} {(l_{ij}-l_0)}^2$, where $\langle ij \rangle$ indicates bonded pairs of nodes and $l_{ij}$ the distance between them. We deform the networks uniaxially under athermal and quasistatic loading conditions, using the FIRE algorithm for the energy minimization~\cite{bitzek2006}. The tolerance $F_{\text{RMS}}$ is typically set at $10^{-5}$ for $L \leq 256$, but it is varied for larger sizes as discussed in the text. In our case, $F_{\text{RMS}}$ is the maximum numerical tolerance for both the system root-mean-square force~\cite{bitzek2006} and the maximum force on each single node. The networks are deformed uniaxially in the $y$-direction by employing a very small strain increment $\Delta \epsilon=0.001$, after which the energy minimization procedure is performed to obtain the equilibrium (off-lattice) node positions. Based on the mechanically equilibrated configuration, we identify the bonds that exceed $\lambda$ (if any) and remove the weakest bond defined as the one that exceeds the threshold the most, i.e. the one with the largest $(l_{ij}-l_0)/\lambda$. In most of the simulations, we additionally allow the strain increment to be reduced by up to two order of magnitude ($\Delta \epsilon=0.00001$) if the most stressed bond is very close to $\lambda$, to ensure that we are in the quasistatic condition and only one bond at a time can be broken. Typically, we use periodic boundary conditions in all directions. However, following Ref.~\cite{zhang2017}, we also perform simulations in which the nodes on the top and bottom boundaries are free to move only in the $x$-direction and move affinely in the $y$-direction, without noticing any appreciable qualitative difference. Furthermore, as detailed below, we also perform simulations by employing different boundary conditions in the $x$-direction. We confirmed that the conclusions of our study are robust with respect to all the variations on the simulation protocol.

\section{Observables and data analysis}
We define the mechanical response of the network by considering the stress along the direction of the deformation. In particular, we use the $yy$-component of the virial stress tensor $\sigma \equiv \sigma_{yy}^{\text{vir}} = \frac{1}{d V} \sum_{\langle ij \rangle} f_{ij,y} \, r_{ij,y}$ , where the sum runs over all the bonded pairs of nodes $\langle ij \rangle$, $\mathbf{f}_{ij}$ is the force acting on node $i$ due to $j$, $\mathbf{r}_{ij}$ is the vector connecting the two nodes, $d$ is the dimensionality of the network and $V$ is its (instantaneous) volume. Using a different definition for the macroscopic stress (e.g. $\sigma_{xx}^{\text{vir}}+\sigma_{yy}^{\text{vir}}$) does not change the conclusions of the paper. A network is defined rigid if $\sigma$ is greater than a small value ($10^{-4}$) and $\epsilon_{\text{R}}$ is the first strain for which this condition is satisfied. $\sigma_{\text{c}}$ is the maximum in the stress response and $\epsilon_{\text{p}}$ the associated strain. The work of extension $W_{\text{f}}$ is calculated as the integral of the stress-strain curve until fracture (network broken in two parts). $\Delta \sigma_{\text{max}}$ is the maximum stress drop and it is calculated following a procedure highlighted in Ref.~\cite{bonfanti2018}. It consists of (i) calculating the derivative of the stress-strain curve; (ii) identifying the strain interval for which the derivative is greater than a certain threshold (here we use zero, but we checked that our results are robust using different thresholds); (iii) for each strain interval, calculate the stress drop as the difference in the stress associated to the extremal values of the interval; (iv) the maximum stress drop is $\Delta \sigma_{\text{max}}$. In most of the cases, this procedure gives exactly the same result as just considering the difference in stress between two consecutive strains; whereas in few cases the values are slightly different but the overall trends are consistent with both definitions.

All quantities are averaged over many configurations to ensure proper statistical sampling of the disorder. Errors are calculated as standard deviation divided by the number of sampled configurations (standard error on the mean) and are shown when they are larger than the symbols displayed in the graphs. For example, for the size scaling analysis for networks close to the isostatic point ($p=0.65$ and $p=0.70$) and for small $\lambda$ we employ more than 2000 configurations for small system sizes ($L \leq 32$); 1000 for $L=50$; 500 for $L=64$; 200 for $L=128,256$; 50 for $L=512$; and $\sim 10$ configurations for larger system sizes ($L=640,768,1024$) that are statistically sufficient due to the self-averaging nature of the fracture process in such large systems. In general, for all networks we average quantities over at least 200 configurations when $L=128$ and a larger number (typically 500) for $L=64$ or smaller sizes. Results for different network geometries are also averaged over approximately 200 configurations for 2D networks and over 20 to 100 configurations for 3D networks, depending on system size.

\section{Additional results}

\subsection{Fixed system size}
\begin{figure*}[h!t]
	\begin{center}
		\includegraphics[width=0.9\textwidth]{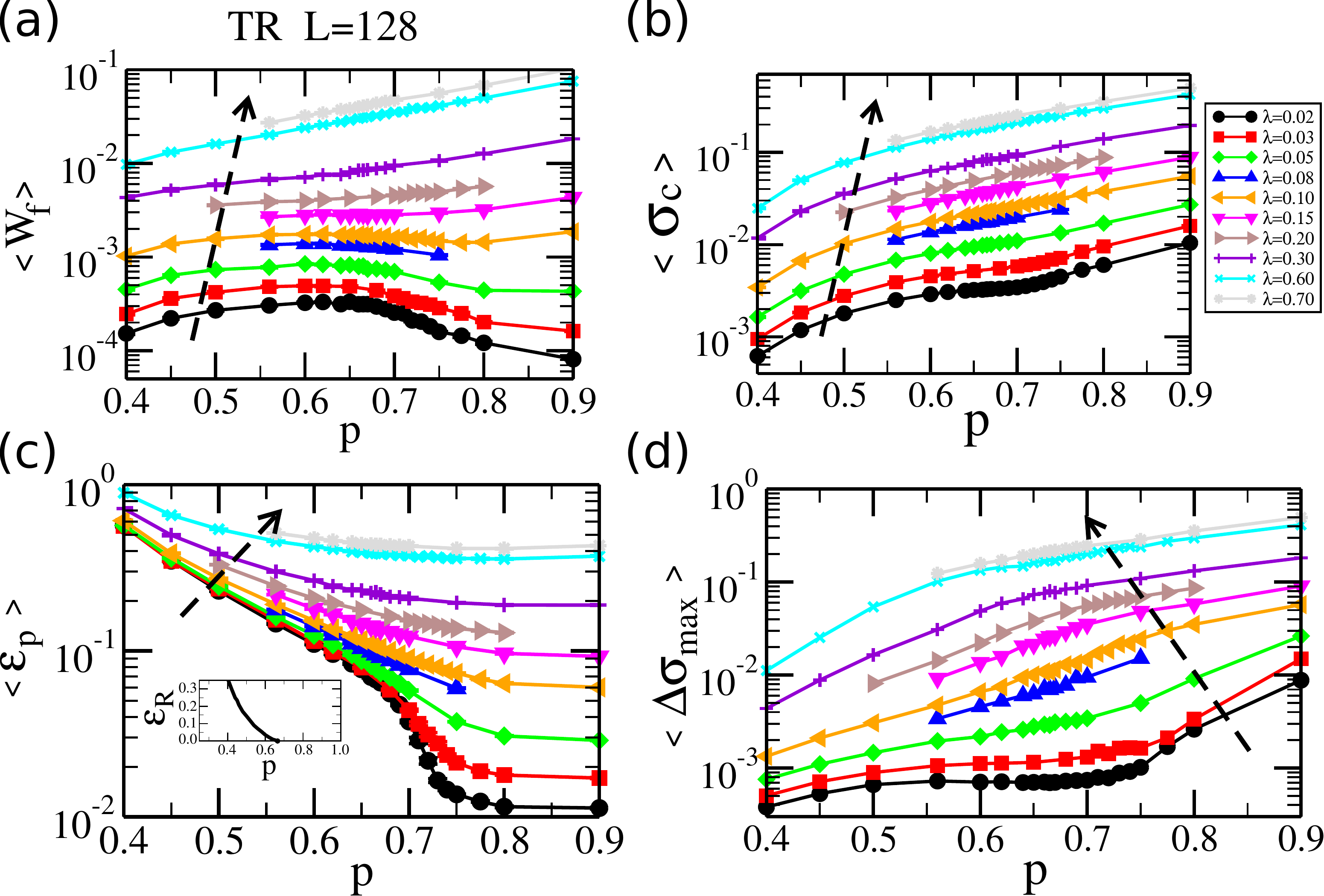}
	\end{center}
	\caption{Additional results for triangular networks with $L=128$ as a function of $p$ and various $\lambda$ (see legend in panel (b)). (a) Work of extension $W_{\text{f}}$, defined as the integral of the stress-strain curve. (b) Maximum stress $\sigma_{\text{c}}$. (c) Strain $\epsilon_{\text{p}}$ corresponding to the stress peak as a function of $p$ for various $\lambda$. Inset: strain $\epsilon_{\text{R}}$ at which the network is rigid upon uniaxial extension as a function of $p$. (d) Maximum stress drop $\Delta \sigma_{\text{max}}$ versus $p$ for different $\lambda$.  }
	\label{SfigA}
\end{figure*}

In Fig.~\ref{SfigA}, some additional results for triangular networks with fixed system size $L=128$ and different $p$ and $\lambda$ are shown. In particular, we note in Fig.~\ref{SfigA}(a) that for fixed $\lambda$ the work of extension $W_f$, i.e. the energy needed to break the material (calculated as the integral of the stress-strain curve), shows a maximum around the isostatic point ($p_{iso}\simeq0.66$) for small $\lambda$, analogously to the trend for $n_f$ (shown in Fig.~1(c) of the main text). This has been observed also in Ref.~\cite{zhang2017} and implies that networks close to the isostatic point can be counterintuitively considered tougher than networks with more bonds (when $\lambda \leq 0.10$). However, as shown in Fig.~\ref{SfigA}(b), when considering the maximum stress $\sigma_{\text{c}}$ as a function of connectivity we observe that there is no maximum and the more intuitive monotonic dependence on $p$ --- the more the bonds, the stronger the material --- is observed. This discrepancy between the dependence on connectivity of two quantities describing resistance to fracture is an effect of the size-dependent brittleness demonstrated in this study: at moderate system size, the networks are ductile and significant energy is dissipated in bond-breaking events occurring after the peak stress. To further quantify the tunability of the fracture process by varying $p$ and $\lambda$ at fixed system size, we show the peak strain $\epsilon_p$ in panel (c) and the maximum stress drop in panel (d).

\subsection{Different topologies}
\begin{figure*}[h!t]
	\begin{center}
		\includegraphics[width=0.75\textwidth]{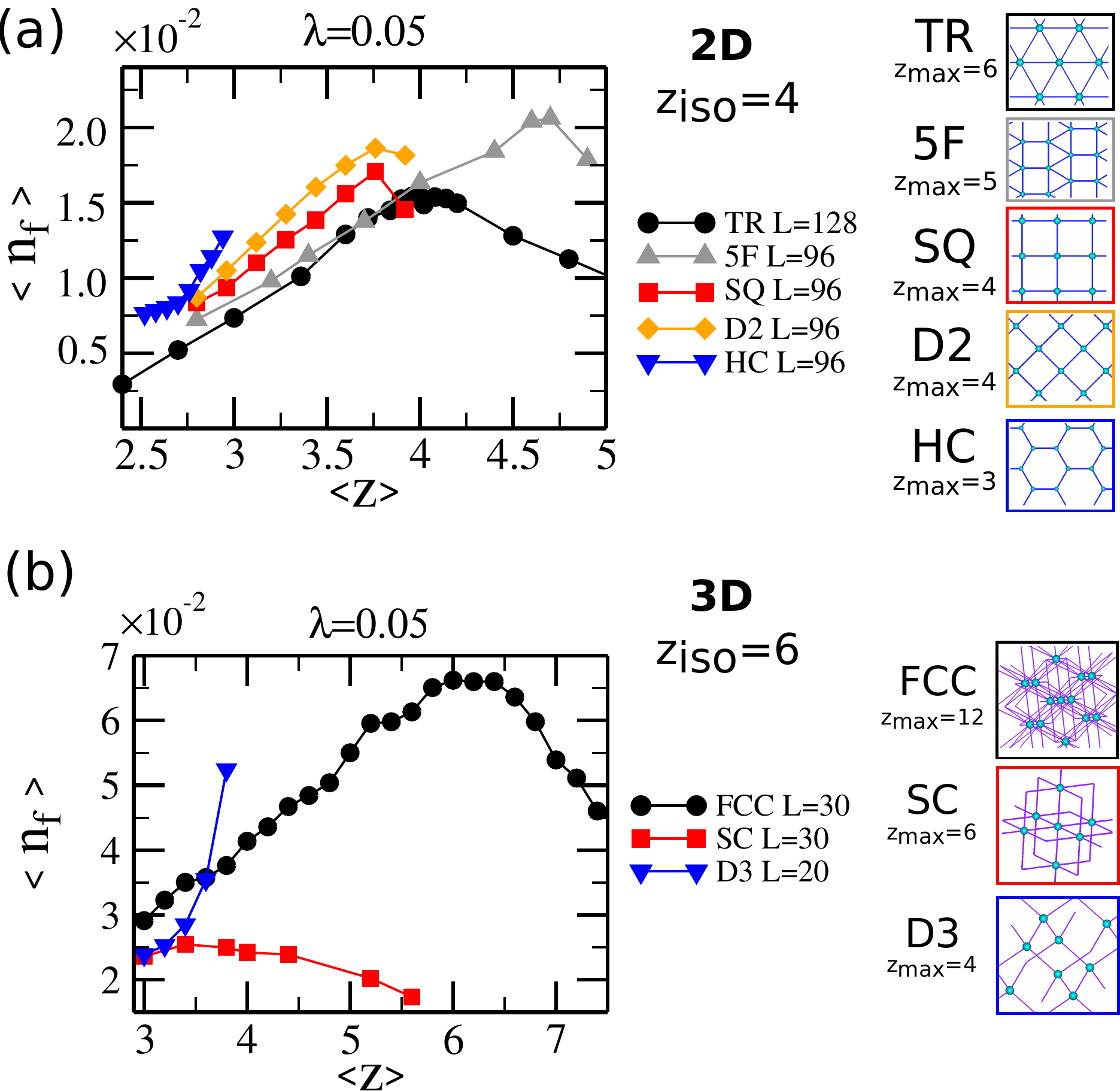}
	\end{center}
	\caption{Fraction of broken bonds as a function of averaged network connectivity $\langle z \rangle$ for (a) 2D and (b) 3D networks with the different architectures shown on the right, same $\lambda$ and similar system size. Maxima are observed close to rigidity points that depend on topology, geometry and dimensionality.}
	\label{SfigB}
\end{figure*}

In Fig.~\ref{SfigB}, we show $n_{\text{f}}$ as a function of the network average connectivity $\langle z \rangle=pz_{\text{max}}$ for different network topology and dimensionality, and small $\lambda$. We observe that the position of the maximum in $n_{\text{f}}$ depends not only on dimensionality but also on topology and orientation. For example, despite the fact that the diluted square (SQ) and the 2D diamond (D2) networks have the same connectivity properties, SQ is more brittle (fewer bonds are broken) because it features more bonds already aligned with the direction of deformation. Analogously, because of geometric effects (a large number of system-spanning bonds at small connectivity) in the honeycomb (HC) and in the 3D diamond (D3) networks, a larger number of bonds is broken when approaching the limit of undiluted lattices. Only for the diluted triangular (TR) and face-centred-cubic (FCC) networks, that are sufficiently isotropic, the maximum of $n_{\text{f}}$ roughly corresponds to the isostatic point $z_{\text{iso}}$. Our results enrich the data reported in Refs.~\cite{zhang2017,driscoll2016,berthier2019}. We can summarize that damage is maximized close to rigidity points, either of isostatic or geometric nature, when $\lambda$ is small.

\subsection{Size scaling}
\begin{figure*}[h!t]
	\begin{center}
		\includegraphics[width=0.9\textwidth]{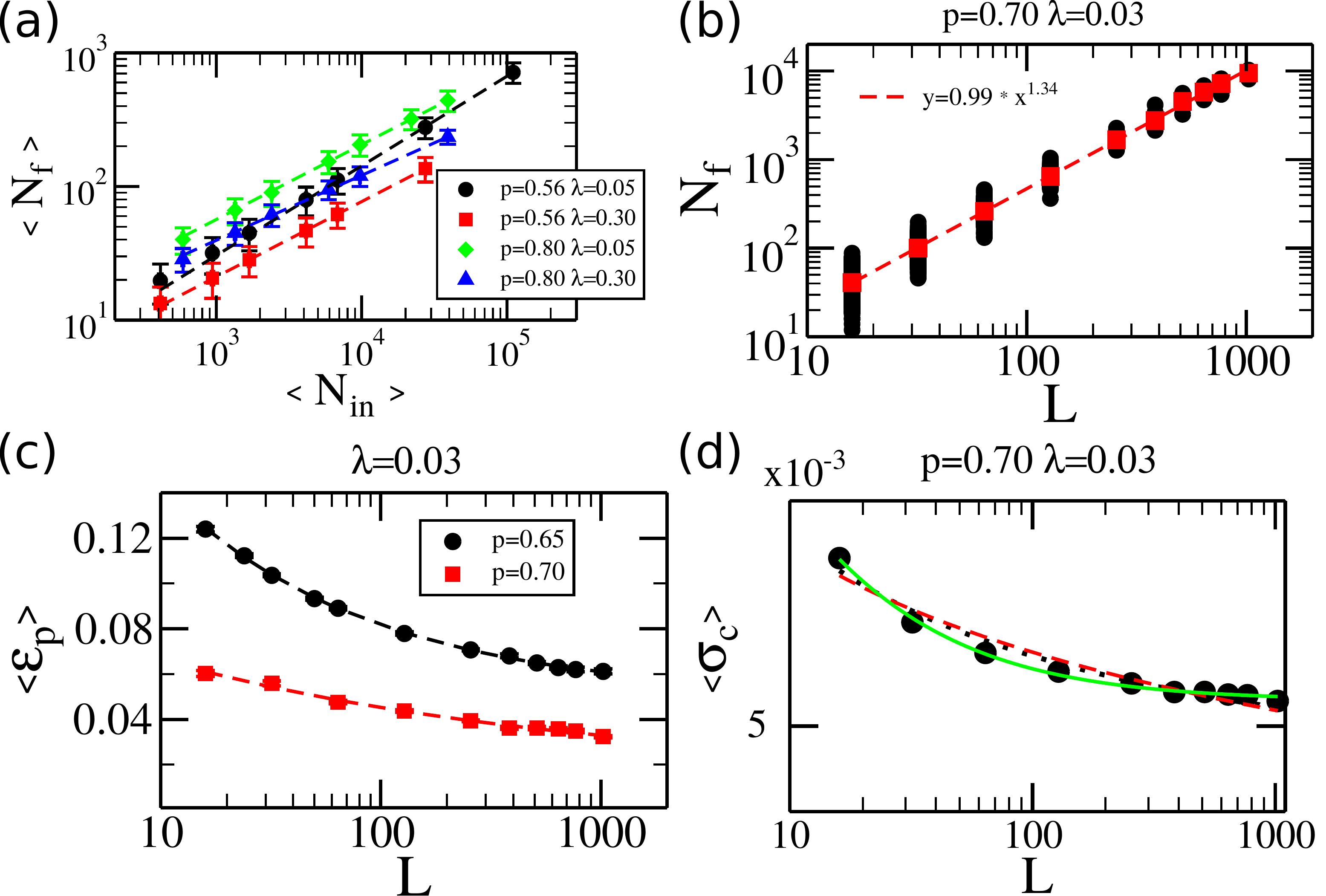}
	\end{center}
	\caption{(a) Number of broken bonds at failure $N_{\text{f}}$ as a function of number of initial intact springs $N_{\text{in}}$ for various combination of $p$ and $\lambda$. Dashed lines are best power-law fits from which the damage fractal dimension $d_{\text{f}}$ is extracted. (b) Size scaling of broken bonds for triangular network close from above to the isostatic point with small $\lambda$. Black circles indicate single-run values, red squares averages and dashed line best power-law fit. (c) Size scaling of strain at peak stress $\epsilon_{\text{p}}$ for networks in the proximity of the isostatic point with small $\lambda=0.03$. Symbols represent averages and lines are best fits. Data are well fitted by a power-law decay with an offset $\epsilon_{\text{p}} = \epsilon_{\text{p}}^{(\infty)} + A/L^{\beta}$, with $\epsilon_{\text{p}}^{(\infty)}$ the strain at peak in the thermodynamic limit. For $p=0.65$, we obtain $\epsilon_{\text{p}}^{(\infty)}=0.050$ and $\beta=0.47$. For $p=0.70$, we obtain $\epsilon_{\text{p}}^{(\infty)}=0.016$ and $\beta=0.24$. (d) Size scaling of the network strength $\sigma_{\text{c}}$ (log-log plot). Symbols correspond to averages, error bars show the standard deviations and lines are the best fits. Three functional forms were used: logarithmic decay (dotted black line), DLB decay (dashed red line), and power-law decay $\sigma_{\text{c}} = \sigma_{\text{c}}^{(\infty)} + A/L^{\beta}$ (green solid line). The latter fits the data the best, we obtain $\sigma_{\text{c}}^{(\infty)}=0.0053$ and $\beta=0.89$. }
	\label{SfigC}
\end{figure*}

In Fig.~\ref{SfigC}, additional results from the size scaling are shown. To extract the damage fractal dimension $d_{\text{f}}$ and generate the plot of Fig.~2(b) of the main text, we simulate at least six different system sizes (typically seven, up to $L=256$) and we fit $N_{\text{f}} \sim N_{\text{in}}^{d_{\text{f}}/2}$ where $N_{\text{in}}$ is the number of intact initial springs. This is analogous to $N_{\text{f}} \sim L^{d_{\text{f}}}$ but sometimes better fits are obtained with the previous expression since the values on the x-axis span more orders of magnitude. As shown in the main text, we notice that the fractal dimension $d_{\text{f}}$ is (much) greater than 1 for a large range of $p$ and $\lambda$. This is a distinct feature of our elastic network model compared to the scalar random fuse models~\cite{alava2006} or models assuming linear elasticity~\cite{nukala2005}, where the number of broken bonds at peak load scales \emph{at most} linearly with the system size. In our case $d_f$ refers to the exponent of the total broken bonds at failure and not to the exponent of broken bonds at peak load. However, since the fracture becomes more brittle (i.e. bonds break only before and at peak load) the exponent for the broken bonds at peak load is actually larger than $d_f$. We also show how the maximum stress $\sigma_{\text{c}}$ and peak strain $\epsilon_p$ decrease with increasing system size. Inspired by previous works~\cite{duxbury1987,manzato2012,roy2017b}, we fitted the decay of $\sigma_{\text{c}}$ with three functional forms: (i) logarithmic decay $\propto 1/\ln(L)$; (ii) Duxbury-Leath-Bale (DLB)~\cite{duxbury1987,manzato2012} $\propto 1/\sqrt{\ln(L^2)}$; (iii) power-law decay $\propto 1/L^{\beta}$. We found that the power-law decay fits our data the best, with an exponent $\beta$ that clearly depends on $p$ and $\lambda$. For example, we obtain $\beta=0.73$ for $p=0.65$, $\lambda=0.03$ (Fig.~2(d) of main text) and $\beta=0.89$ for $p=0.70$, $\lambda=0.03$ (Fig.~\ref{SfigC}(d)). Differently from previous models where it is assumed that the material strength (i.e. $\sigma_c$) is zero in the thermodynamic limit~\cite{alava2006}, in our case we have an offset value corresponding to a finite network strength $\sigma_c^{(\infty)} \neq 0$ that depends on $p$ and $\lambda$. Our data suggest that the spring network model studied here is qualitatively different from scalar models that have been so far considered as the golden standard for statistical physics studies of fracture.

\subsection{Universality of fracture abruptness trend}
\begin{figure*}[h!t]
	\begin{center}
		\includegraphics[width=0.8\textwidth]{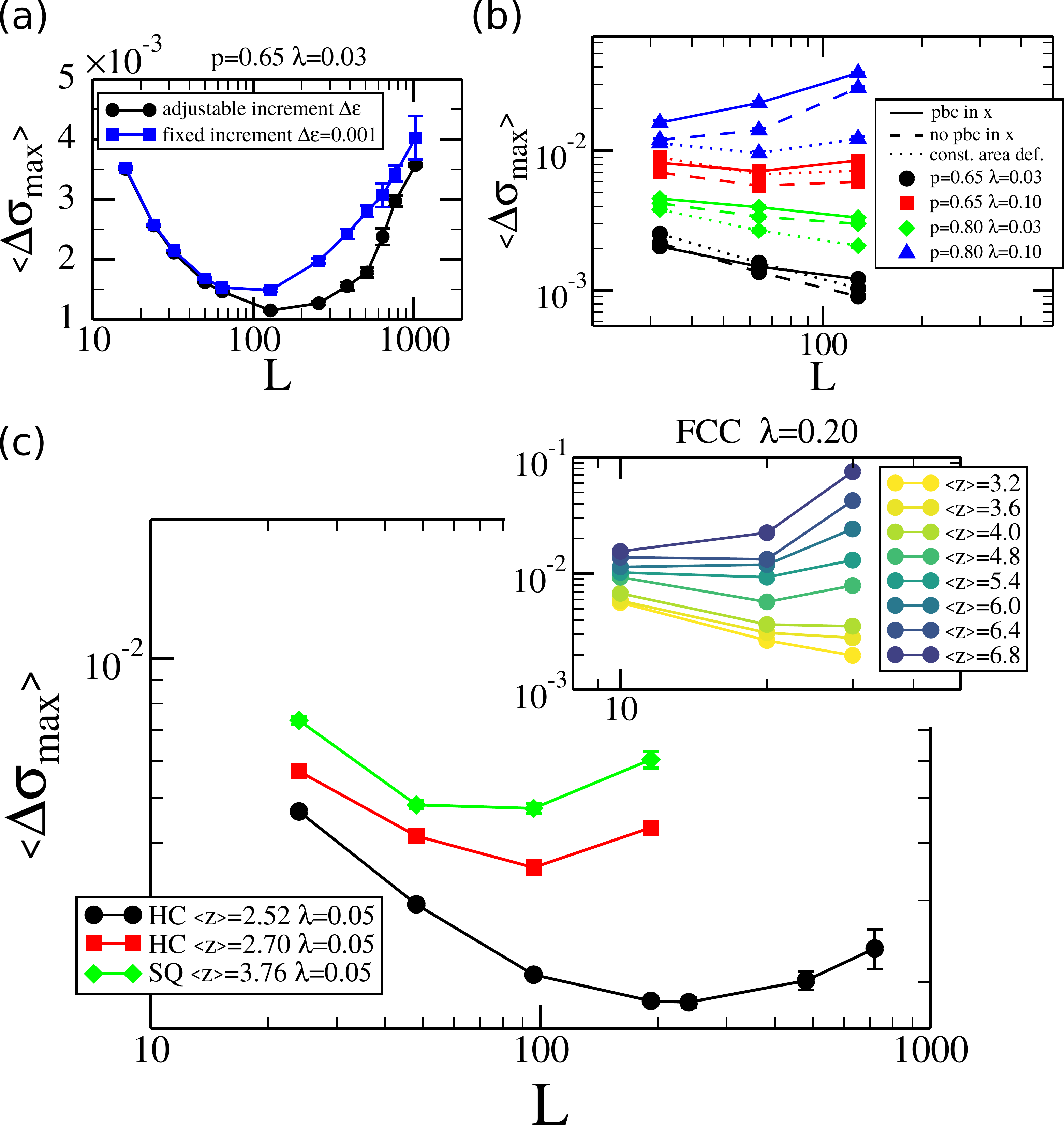}
	\end{center}
	\caption{(a) Size scaling of maximum stress drop for two different simulation protocols. Black circles represent data from simulations where the strain step can be adjusted (reduced) to smaller steps when there are bonds very close to their thresholds (see section on simulation methods); whereas blue squares are data from simulations where the strain step is kept fixed at a small value. (b) Size scaling of maximum stress drop for four networks with $p$ and $\lambda$ as indicated in the legend, and three different simulation protocols for which periodic boundary conditions in the $x$ direction (normal to the deformation direction) are present (solid lines) or not (dashed lines), or for which the $x$ direction is adjusted to keep the system area constant (dotted lines). (c) Size scaling of fracture abruptness for diluted 2D honeycomb (HC), 2D square (SQ) and 3D FCC (inset) networks with averaged connectivity $\langle z \rangle$ and $\lambda$ as indicated.}
	\label{SfigD}
\end{figure*}

\begin{figure*}[h!t]
	\begin{center}
		\includegraphics[width=0.9\textwidth]{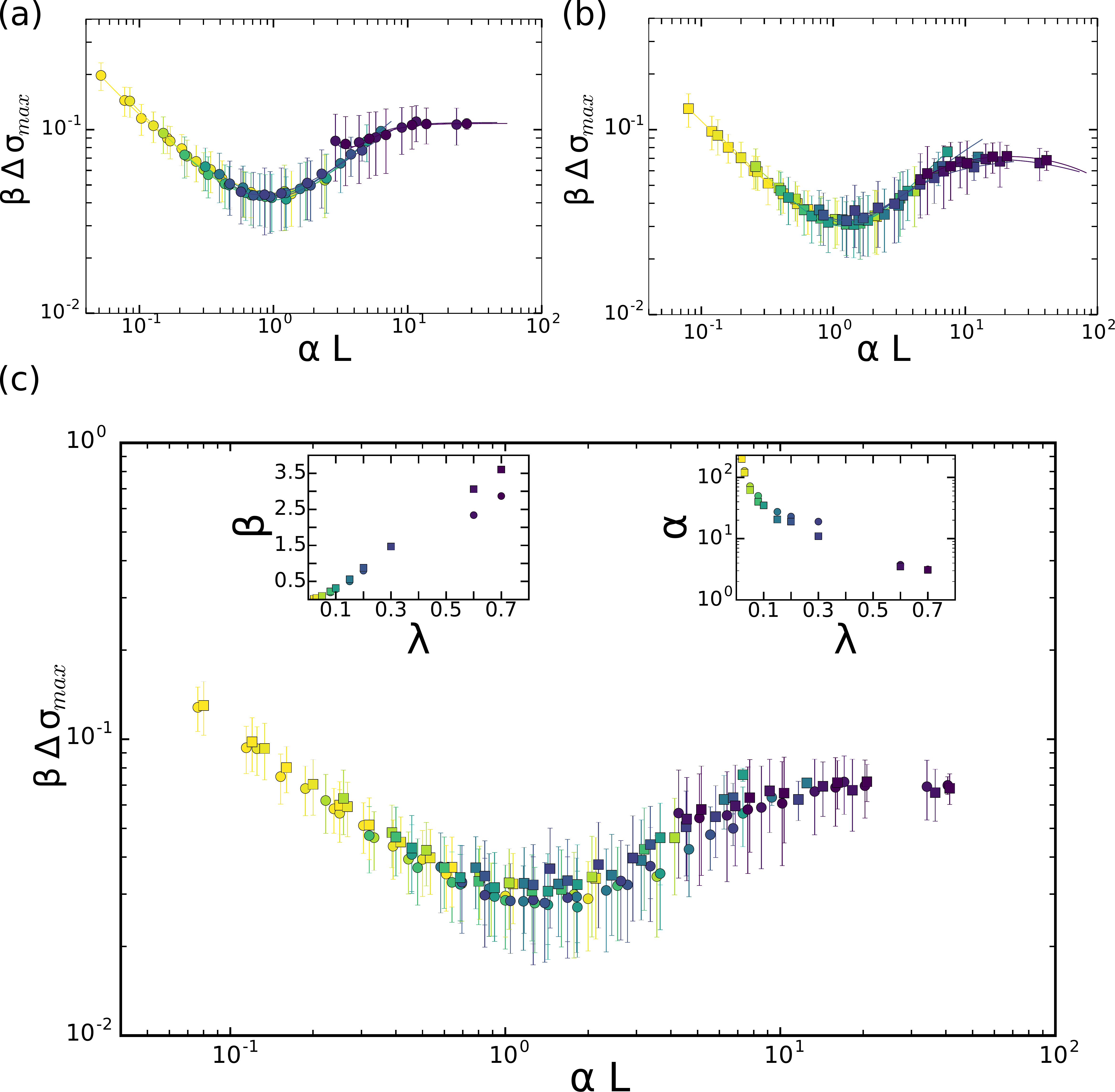}
	\end{center}
	\caption{Rescaled maximum stress drop $\Delta \sigma_{\text{max}}$ as a function of rescaled system size $L$ for (a) $p=0.65$ (see also Fig.~3(d) of main text) and (b) $p=0.70$, color-coded according to $\lambda$. Symbols represent averages and error bars are standard deviations. (c) Both data sets combined (circles for $p=0.65$, squares for $p=0.70$). The values used for the rescaling parameters $\alpha$ and $\beta$ are shown in the insets as a function of $\lambda$. $\beta$ grows (more than linearly) with increasing $\lambda$. $\alpha$ is related to the transition system size $L^*$ discussed in the main text and decreases by two order of magnitudes in the simulated range of $\lambda$. }
	\label{SfigE}
\end{figure*}

In Figs.~\ref{SfigD} and~\ref{SfigE}, we show that the non-monotonic size-dependence of fracture abruptness $\Delta \sigma_{\text{max}}$ is universal with respect to network topology and robust with respect to the simulation procedure. In Fig.~\ref{SfigD}(a), we compare the trend for the same set of initial configurations simulated with a protocol where the small strain increment is kept fixed at $\Delta \epsilon=0.001$ and one where it can be adjusted to much smaller values (up to $\Delta \epsilon=0.00001$) if the most stressed bond is very close to $\lambda$. A minimum is observed in both cases for similar values of $L$. A slightly smaller $\Delta \sigma_{\text{max}}$ is observed for large $L$ when $\Delta \epsilon$ can be reduced by two orders of magnitude. Next, we check if the choice of boundary conditions in the $x$-direction, transverse to the deformation direction, has some qualitative effect on the fracture abruptness. We tested three different boundary conditions: (i) periodic (mimicking an infinitely large system); (ii) free, i.e. without bonds crossing the $x$-direction (the system is finite in that direction); (iii) with periodic bonds and with adjustable size in the $x$-direction such that the area is kept constant during the uniaxial deformation. As shown in Fig.~\ref{SfigD}(b), the qualitative trends are robust and independent of the fundamentally different boundary conditions employed (the first one was chosen for the rest of our study since it seems the most computationally efficient). Furthermore, we confirm that the non-monotonic trend persists when changing network topology. As already highlighted in the main text, the minimum and the maximum of $\Delta \sigma_{\text{max}}$ as a function of $L$ depend on network properties. This can be appreciated in Fig.~\ref{SfigD}(c) where we report selected cases for various 2D and 3D networks. Finally, preliminary results confirm the non-monotonic trend also for shear deformation, supporting the conclusion that it is a universal size-dependent behavior of diluted networks. As shown in Fig.~\ref{SfigE}, we also succeeded in manually collapsing the curves of $\Delta \sigma_{\text{max}}$ obtained for networks close to (just below and slightly above) the isostatic point and various $\lambda$ into a single master curve. The rescaling $L \rightarrow \alpha L$ and $\Delta \sigma_{\text{max}} \rightarrow \beta \Delta \sigma_{\text{max}}$ was performed by imposing that the different curves approximately share the same x-value for the minimum. We obtained values for $\alpha$ that are inversely proportional to $\lambda$, suggesting that the ``critical'' system size $L^* \sim 1/\alpha$ at which brittleness kicks in decreases by making the springs stronger. On the other hand, $\beta$ increases more than linearly with $\lambda$, consistently with the idea that due to inhomogeneous stress distribution, the increase of the springs strength has non-linear consequences on the overall network.

\subsection{Effect of tolerance in energy minimization}
\begin{figure*}[h!t]
	\begin{center}
		\includegraphics[width=0.9\textwidth]{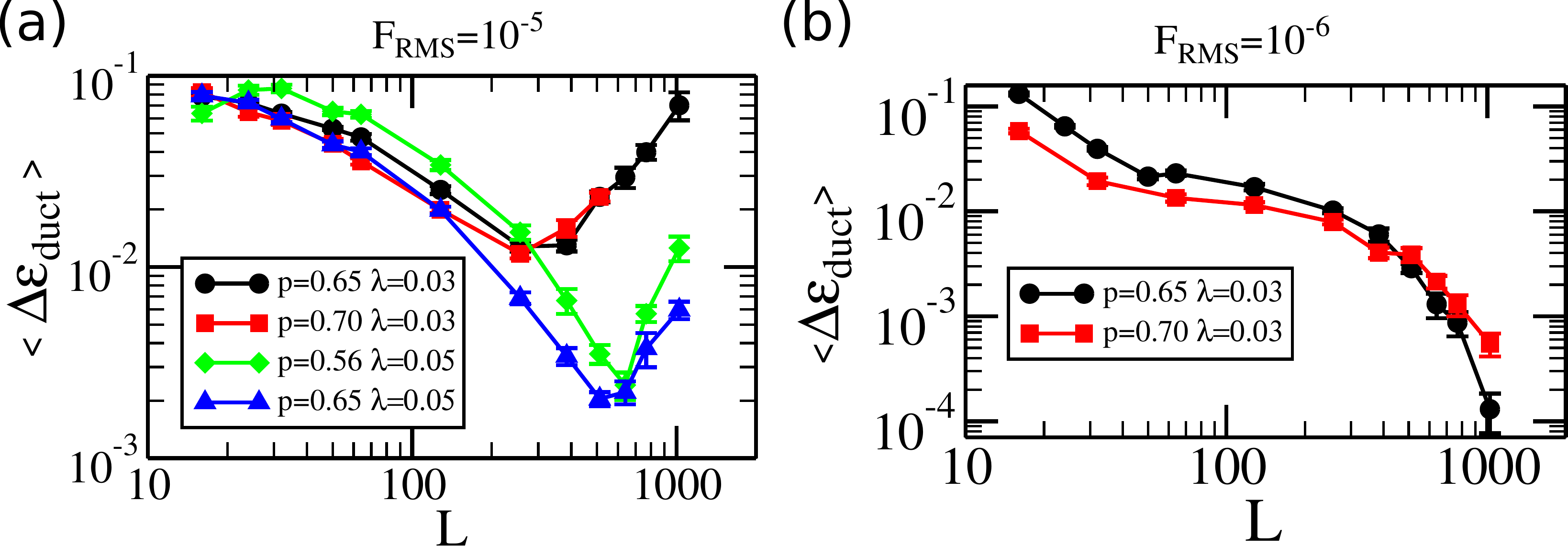}
	\end{center}
	\caption{Size scaling of the ductility interval $\Delta \epsilon_{\text{duct}} = \epsilon_{\Delta\sigma_{\text{max}}} - \epsilon_{\text{p}}$, where $\epsilon_{\Delta\sigma_{\text{max}}}$ is the strain at which the maximum stress drop is observed (see also Fig.~4 of the main text), for simulations with (a) $F_{\text{RMS}}=10^{-5}$ and (b) $F_{\text{RMS}}=10^{-6}$ for different $p$, $\lambda$ as indicated in the legend. For small $F_{\text{RMS}}$ (panel (b)), it is evident that fracture becomes more brittle upon increasing system size. For larger $F_{\text{RMS}}$ (panel (a)), that are nevertheless three orders of magnitude smaller than $\lambda$, a non-monotonic trend is observed and networks with large $L$ reacquire ductility. At a first glance, the minima in $\Delta \epsilon_{\text{duct}}$ seem to roughly correspond with the minima in $\Delta \sigma_{\text{max}}$ (see Fig.~3(a) of main text), but further studies are needed to confirm this correspondence. }
	\label{SfigF}
\end{figure*}

Finally, in Fig.~\ref{SfigF} we report the size scaling of the ductility interval $\Delta \epsilon_{\text{duct}}$, defined as the strain difference between $\epsilon_p$ and the strain corresponding to the maximum stress drop, for two different values of the tolerance $F_{\text{RMS}}$ and various networks. Brittle fracture (vanishing $\Delta \epsilon_{\text{duct}}$) is observed in the athermal limit, whereas a non-monotonic size-dependence of the ductility interval is observed for larger $F_{\text{RMS}}$. The system size at which ductility is increasing again (the minimum in $\Delta \epsilon_{\text{duct}}$ as a function of $L$) does strongly depend on $\lambda$ and therefore seems to be coupled with the network rigidity. If a variation of the tolerance in the energy minimization protocol $F_{RMS}$ is interpreted as a variation of the thermal effects in the system (that would keep the system not exactly at the energy minimum), these results might suggest that by increasing temperature it is possible to observe a more ductile fracture occurring via many large cracks. Future research should further investigate the intriguing possibility that the minimum temperature at which such ductile behavior is first observed, and the material rigidity are strongly coupled. However, to precisely assess the role of temperature, more traditional simulation methods for thermal systems should be employed.

\clearpage

\end{document}